\documentclass[prl,twocolumn]{revtex4}
\usepackage{graphicx}
\usepackage{rotating}
\usepackage{setspace}

\begin{document}
\title{Single Crystal Study of Competing Rhombohedral and Monoclinic Order in Lead Zirconate Titanate}
\author{D. Phelan$^{1}$, X. Long$^{2}$, Y. Xie$^{2}$, Z.-G. Ye$^{2}$, A. M. Glazer$^{3}$, H. Yokota$^{4}$, P. A. Thomas$^{5}$, P. M. Gehring$^{1}$}
\affiliation{
$^{1}$NIST Center for Neutron Research, National Institute of Standards and Technology, Gaithersburg, Maryland 20899, USA\\
$^{2}$Department of Chemistry, Simon Fraser University, Burnaby, BC, V5A 1S6, Canada\\
$^{3}$Clarendon Laboratory, Department of Physics, University of Oxford, Parks Road, Oxford OX1 3PU, United Kingdom\\
$^{4}$Department of Bioengineering, University of Tokyo, Tokyo 113-8656, Japan\\
$^{5}$Department of Physics, University of Warwick, Coventry CV4 7AL, United Kingdom\\
}

\begin{abstract}
Neutron diffraction data obtained on single crystals of PbZr$_{1-x}$Ti$_x$O$_3$ with x = 0.325 and x = 0.460,
which lie on the pseudorhombohedral side of the morphotropic phase boundary, suggest a coexistence of
rhombohedral (\textit{R3m}/\textit{R3c}) and monoclinic (\textit{Cm}) domains and that monoclinic order is enhanced by
Ti substitution. A monoclinic phase with a doubled unit cell (\textit{Cc}) is ruled out as the ground state.
\end{abstract}
\maketitle

A hallmark of highly piezoelectric perovskite materials,
such as PbZr$_{1-x}$Ti$_x$O$_3$ (PZT), is a morphotropic phase
boundary (MPB) that segregates two compositional regimes
into different ferroelectric states and pinpoints the
range of substitutions for which the piezoelectric response
is optimal. For PZT (and other Pb-based materials), the
MPB ostensibly divides regions of Ti-poor rhombohedral
(\textit{R3m}) and Ti-rich tetragonal (\textit{P4mm}) order. The discovery
of an intermediate monoclinic phase within a narrow compositional
range near the MPB [1,2] provided a mechanism
whereby PZT could transform from R3m to P4mm via a
common subgroup, Cm. This led to the idea that enhanced
piezoelectricity at the MPB arises from the freedom of the
Pb atom to move within the mirror plane of the monoclinic
phase, thus allowing the electric polarization to rotate
[3,4]. Alternatively, the adaptive-phases model, proposes
that the lower symmetry allows for complex domain structures
in which the net polarization can be made to change
direction via domain populations [5]. These mechanisms
remain controversial as how the phase symmetry is transformed
across the phase diagram remains unclarified.

The most recent measurements identify no clear phase
boundary between rhombohedral and monoclinic phases
[6]. Instead the anisotropic Pb displacement ellipsoid is
highly flattened perpendicular to the polar displacement
direction, which led to the suggestion that some kind of
disorder [7,8] or local monoclinic displacements [9] are
present in the rhombohedral phases. From this followed the
idea that the local symmetries are monoclinic, but that the
coherence lengths are limited to just a few unit cells so that
diffraction methods yield only rhombohedral symmetries
for Ti-poor compositions [10]. While this model is able to
explain the lack of an R-M phase boundary, two different
explanations were recently put forth. The first, based on
refinements of synchrotron data, replaces the rhombohedral
phases with a high-temperature monoclinic \textit{Cm} and a
low-temperature monoclinic \textit{Cc} phase (the \textit{Cc} phase
superimposes antiferrodistortive displacements with the
ferroelectric displacements of \textit{Cm}, causing the unit cell
to double as depicted in Ref. [11]) [11,12]. The second,
based on a separate set of neutron diffraction refinements
favors a coexistence of rhombohedral \textit{R3c} and monoclinic
\textit{Cm} (but not \textit{Cc}) phases at room temperature in the Ti-poor
region [6]. The present results resolve this issue
conclusively.

A hindrance to precise symmetry determination of PZT
has been the difficulty to grow single crystals. Therefore,
structural studies near the MPB have been limited to powders
and ceramics. A major problem with Rietveld refinement
of highly pseudosymmetric polycrystalline materials
is that many models refine to similar agreement factors,
which makes it nearly impossible to choose between them.
Very recently, single crystals were grown at Simon Fraser
University using a top-seeded solution growth technique.
The crystals measured here have compositions of x =
0.325 (dimensions 3.3 mm $\times$ 2.1 mm $\times$ 1.1 mm) and x =
0.460 (3.0 mm $\times$ 2.7 mm $\times$ 0.6 mm). The measurements
were performed at the NIST Center for Neutron Research
(NCNR) on thermal (BT-9) and cold (SPINS) triple-axis
spectrometers. Various instrumental configurations were
employed and are given in the figure captions. The samples
were measured under vacuum in a closed-cycle helium
refrigerator. Miller indices are expressed with respect to
the aristotype cubic cell with a lattice parameter a $\approx$ 4.1 \AA.

Structural transitions in each crystal were characterized
by measurements of the 200 Bragg reflection as a function
of temperature (see Fig. 1). For x = 0.325, two phase
transitions were identified corresponding to paraelectric-ferroelectric
and ferroelectric-ferroelectric transitions. One
occurs near 370 K and is marked by a change in the slope of
the temperature dependence of the rocking curve width, a
strong shift in the peak position of the rocking scan, and a
change from positive to negative volume thermal expansion.
A second near 590 K is evident from the large change in intensity resulting from a release of secondary extinction,
a leveling out of the width of the rocking curve, and a return
to positive volume thermal expansion. Similar measurements
for x = 0.460 revealed more complicated behavior.
Possible phase transitions are seen at 220, 410, 480, 540,
and 650 K. Between 480 and 540 K the volume thermal
expansion is $\approx$ 9.1 $\times$ 10$^{-5}$ 1/K, or roughly 1 order of
magnitude larger than normally observed in oxides; above
540 K a steep negative volume thermal expansion
($\approx$ -4.9 $\times$ 10$^{-5}$ 1/K) is observed. A huge release of
extinction occurs upon cooling below the cubic phase.
The large number of anomalies is consistent with the fact
that the crystal has a composition very close to the MPB.
A similarly complex series of phase transitions has been
suggested [13] based on powder neutron data [14].

One approach to identify the structural symmetry is to
look for superlattice peaks that appear when the primitive
unit cell is doubled. The \textit{R3c} phase generates superlattice
reflections that result from correlated rotations of the oxygen
octahedra, but only at the $R_2$ points $\frac{h}{2}\frac{\overline{k}}{2}\frac{k}{2}$,
where h and k are odd and h $\neq$ k (for twinned crystals, $\frac{h}{2}\frac{\overline{k}}{2}\frac{k}{2}$ and
$\frac{h}{2}\frac{k}{2}\frac{k}{2}$ are not distinct).
This is the only type of superlattice peak that
has been observed in x-ray and neutron powder diffraction
measurements [6,15,16]. Other superlattice reflections
have been observed using electron diffraction methods
[15,17-20]; however, these have been argued to result
from surface effects or local inhomogeneities, with only
$R_2$ attributed to the bulk [15]. On the other hand, a phase
with \textit{Cc} symmetry [11,12] would generate weak superlattice
reflections at the $R_1$ points $\frac{h}{2}\frac{h}{2}\frac{h}{2}$, where h is odd. 
We looked for superlattice reflections in the x = 0.325 crystal
in the (hk0) and (hkk) scattering planes at 35 K (see Fig. 2).
No peaks were found at the $X$ points $\frac{1}{2}$00 and $\frac{3}{2}$00, or at the
$M$ points $\frac{1}{2}\frac{1}{2}$0 and $\frac{3}{2}\frac{1}{2}$0, however, several $R_2$ peaks were
observed. The order parameter measurement shown in
Fig. 2(f) indicates that the $R_2$ peaks vanish at the
low-temperature phase transition identified in Fig. 1(a)
($\approx$ 370 K). Weaker $R_2$ peaks were observed in the x =
0.460 crystal that vanish above the lowest transition temperature
identified in Fig. 1(b) ($\approx$ 220 K). We also observed
superlattice peaks at the $R_1$ points $\frac{1}{2}\frac{1}{2}\frac{1}{2}$ and $\frac{3}{2}\frac{3}{2}\frac{3}{2}$,
which showed temperature dependences similar to those of
the $R_2$ points. Because these peaks could be caused by
double scattering as for SrTiO$_3$ [21], we performed a standard
test in which the peak intensity was measured as a
function of the neutron wavelength. Figure 2(e) shows
measurements performed with cold neutrons at two different
wavelengths. A very strong $R_1$ peak was observed
at $\frac{1}{2}\frac{1}{2}\frac{1}{2}$ when $\lambda$ = 4.045 \AA; however, when the wavelength
was increased to 5.222 \AA, so that all $R_2$ reflections fell
outside the Ewald sphere, this $R_1$ peak completely disappeared.
Hence, we can definitively conclude that the observed
$R_1$ peak results from double scattering. We have
determined with 99\% statistical confidence that the maximum
ratio of the peak intensity $\frac{1}{2}\frac{1}{2}\frac{1}{2}$ compared with that of
100 is 0.002\% (x = 0.325) and 0.003\% (x = 0.460). These
are extremely severe limits: for the refinement listed in
Ref. [22], which is for a sample with x = 0.48 at 10 K,
the calculated intensity ratio is 0.13\% (nearly 40 times
larger than our limit for x = 0.46), while refinement for a
sample with x = 0:30 at 300 K [23] gives a ratio of 3.8\%
(more than 1000 times larger). Therefore, our superlattice
survey is consistent with an \textit{R3c} ground state and effectively
rules out the presence of any phase, such as \textit{Cc} or even
triclinic symmetry, that would generate $R_1$ superlattice
peaks for either sample.

Symmetries were further studied by high resolution radial
measurements of several Bragg reflections, which is
possible because the crystals are multidomain. The 200
reflection can be used to distinguish between \textit{Cm} and
\textit{R3m}/\textit{R3c} phases because it splits into a doublet under
\textit{Cm} but remains a singlet for \textit{R3m}/\textit{R3c}. We achieved
extraordinary wave-vector resolution for this peak by employing
a perfect single crystal of Ge as analyzer and
exploiting the fact that the Ge 220 reflection d spacing
almost exactly matches that of the PZT 200 Bragg peak.
A peculiarity of this special setup is that the radial linewidth
becomes coupled to the sample mosaicity [24], particularly
in the regime where the mosaic spread is less than 20'. As
shown in Fig. 3(a), a narrow radial linewidth was observed
at 200 in the cubic phase (640 K) for x = 0.325 that is
nonetheless significantly broader than that measured on a
single crystal of SrTiO$_3$. The extra linewidth is well
accounted for by the small, but non-negligible mosaic width
of the x = 0.325 PZT crystal (11') that was measured under
the same conditions in a rocking scan, and which is taken
into account in the resolution calculation. From Fig. 1(a) we
know that the mosaic width increases as the temperature is
lowered; thus the 200 radial linewidth will necessarily
broaden as the temperature decreases. Indeed, the 200 radial
linewidth at 450K is 12\% broader than that at 640 K, but this
agrees almost perfectly with the resolution calculation for
the broadened mosaic width; hence the radial linewidth of
200 at 450 K is consistent with a resolution-limited peak as
would be expected for \textit{R3m}. However, at 40 K the peak has
broadened by an additional 10\% from its value at 450 K,
while our resolution calculation predicts a change of only
2\%. Given the larger intrinsic mosaic width of the x =
0.460 its 200 reflection was measured using very tight
beam collimations and a conventional PG analyzer, for
which the resolution is essentially decoupled from the
mosaic width. These data, shown in Fig. 3(b), reveal a
broadening of 200 as the temperature is lowered that is
even larger than that for x = 0.325.

The broadened 200 Bragg peaks could result from microscopic
strain, finite domain sizes, or a very small \textit{Cm}
distortion [6]. Microscopic strain and unresolved monoclinic
distortions have the same Q dependence and are
impossible to distinguish in our measurements. However,
the intrinsically broad 200 linewidths are consistent with
recent neutron powder diffraction studies that indicate
that the best refinements are obtained with a model of
coexisting \textit{R3c} and \textit{Cm} phases [6]. It is difficult to extract
quantitative information about the \textit{Cm} phase because the
200 Bragg peak is only slightly broader than the instrumental
resolution and can be fit in many different ways.
However, we can extract important information about the
\textit{R3c} phase because it generates a superlattice peak that
depends only on the rhombohedral structural correlations.
To this end we measured radial scans through the $R_2$
superlattice peak for both compositions with tight beam
collimations at low temperature, as shown in Fig. 4. The
two crystals exhibit markedly different linewidths: the
peak for x = 0.325 is narrow and close to the resolution
limit whereas that for x = 0.460 is severely broadened. A
superposition of two resolution-limited Gaussians cannot
reproduce this broadening. Instead, the broadening implies
the presence of either finite regions of correlated octahedral
rotations or microstrain. Generally, there is no microstrain
associated with oxygen octahedral rotations. Thus,
the most likely explanation for the change in linewidth is
that the coherence length of the rotations is diminished.
This is an intriguing result, especially when considered in
conjunction with the broadened 200 Bragg peaks, because
it suggests that the structural correlations in PZT gradually
evolve with Ti content from being predominantly longrange
rhombohedral for compositions far from the MPB to
being predominantly long-range monoclinic for compositions
close to the MPB.

The radial line shapes of 200, the appearance of superlattice
peaks at $R_2$ points at low temperature, and the
absence of superlattice peaks at R1 points are consistent
with a purely R3m high-temperature ferroelectric phase for
x = 0.325 and coexisting (R3m) and (Cm) phases for x =
0.460, and coexisting R3c and Cm ground-state ferroelectric
phases for both compositions. In addition, the marked
broadening of the superlattice peak at $\frac{3}{2}\frac{1}{2}\frac{1}{2}$, which is
generated by \textit{R3c} symmetry, suggests a scenario of competing
rhombohedral and monoclinic order in highly piezoelectric
compositions in which the rhombohedral
correlations become increasingly short-range upon approaching
the MPB. This picture lends support to the
recent theoretical ideas that the tendency to form macroscopic
monoclinic phases facilitates the mechanism of
polarization rotation [25], achieved either by having the
freedom to change the direction of Pb displacements
within monoclinic unit cells or through the change in the
population of twinned monoclinic components under an
applied stress or electric field. The elucidation of the
correct symmetries of PZT close to the MPB is critical to
the theory of the origin of the high piezoelectricity in this
and other materials, since it affects the discussion of the
numbers and types of microdomains expected as well as
the microscopic mechanisms for polarization rotation.

A. M. G. and P. A. T. are grateful for funding from the
Engineering and Physical Sciences Research Council
(EPSRC) and from the National Science Foundation
(NSF). X. L., Y. X, and Z.-G.Y. acknowledge support from
the U.S. Office of Naval Research (Grant No. N00014-06-1-
0166). This work utilized facilities supported in part by the
NSF under Agreement No. DMR-0944772.

\textbf{Figure Captions:}\newline
\newline
\textbf{Fig. 1:}\newline
Temperature dependence of the 200
Bragg reflection for (a) x = 0.325 and (b) x = 0.460. Shown
are the integrated intensity, full-width-at-half-maximum
(FWHM), and peak position ($\omega_s$) of Gaussian fits to rocking
(transverse) scans through the Bragg peak. Also shown is the
average unit cell volume (aristotype) determined from Gaussian
fits to radial scans through the Bragg peak. Apparent phase
transition temperatures are marked by dashed lines. Error bars
correspond to uncertainties of 1$\sigma$ in the fitted parameters.
Measurements were performed on BT-9 with horizontal beam
collimations of 40'-47'-40'-80' and a single PG filter.
\newline
\newline
\textbf{Fig. 2:}\newline
(a)-(c) Radial scans through the $X$, $M_1$, and $M_2$ points reveal no superlattice reflections for x = 0.325 at 35 K.
(d) Scans through one $R_2$ point at various temperatures for x = 0.325. (e) Scans performed using two wavelengths through an $R_1$ point
for x = 0.325. (f) The order parameters of the superlattice phases are given by the peak intensity of the $\frac{5}{2}\frac{1}{2}\frac{1}{2}$ reflection. Error bars
represent an uncertainty of 1$\sigma$ in the measured intensities. Scans in (a)-(d) and (f) were made using collimations of 40'-47'-40'-80',
3 PG filters, and $\lambda$ = 2.359 \AA\ on BT-9. The scans in (e) were made on SPINS with collimations of 80'-80' and two Be filters.
\newline
\newline
\textbf{Fig. 3:}\newline
(a) Radial scans through the 200 Bragg
reflection of the x = 0.325 crystal at selected temperatures. The
measurements were performed on BT-9 with $\lambda$ = 2.359 \AA,
collimations of 15'-47'-10'-20', a single PG filter, and the Ge
220 analyzer. Also shown is a measurement of single crystal
SrTiO$_3$ (data from Ref. [24] represented by green triangles)
obtained with collimations of 10'-40'-20'-40'. (b) Radial scans
through 200 for the x = 0.460 crystal using the PG 002 analyzer,
3 PG filters, and collimations of 15'-10'-10'-10'.
\newline
\newline
\textbf{Fig. 4:}\newline
High resolution radial scans performed
through the $R_2$ point on SPINS using collimations of G (39')-
40'-10'-20', two Be filters, and with $\lambda$ = 4.045 \AA.\newline
\newline
\clearpage
\includegraphics[scale=0.9]{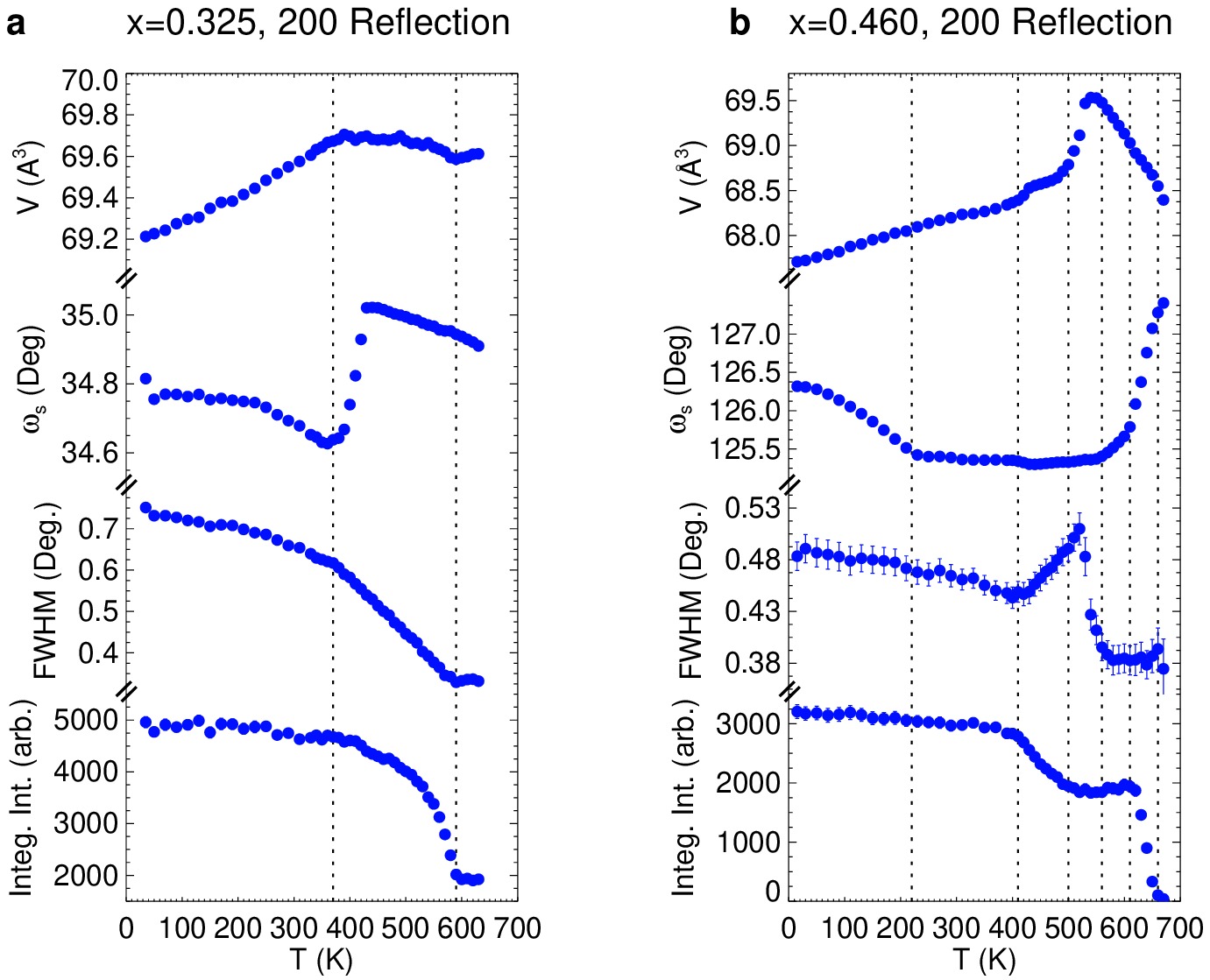}
\clearpage
\includegraphics[scale=0.9]{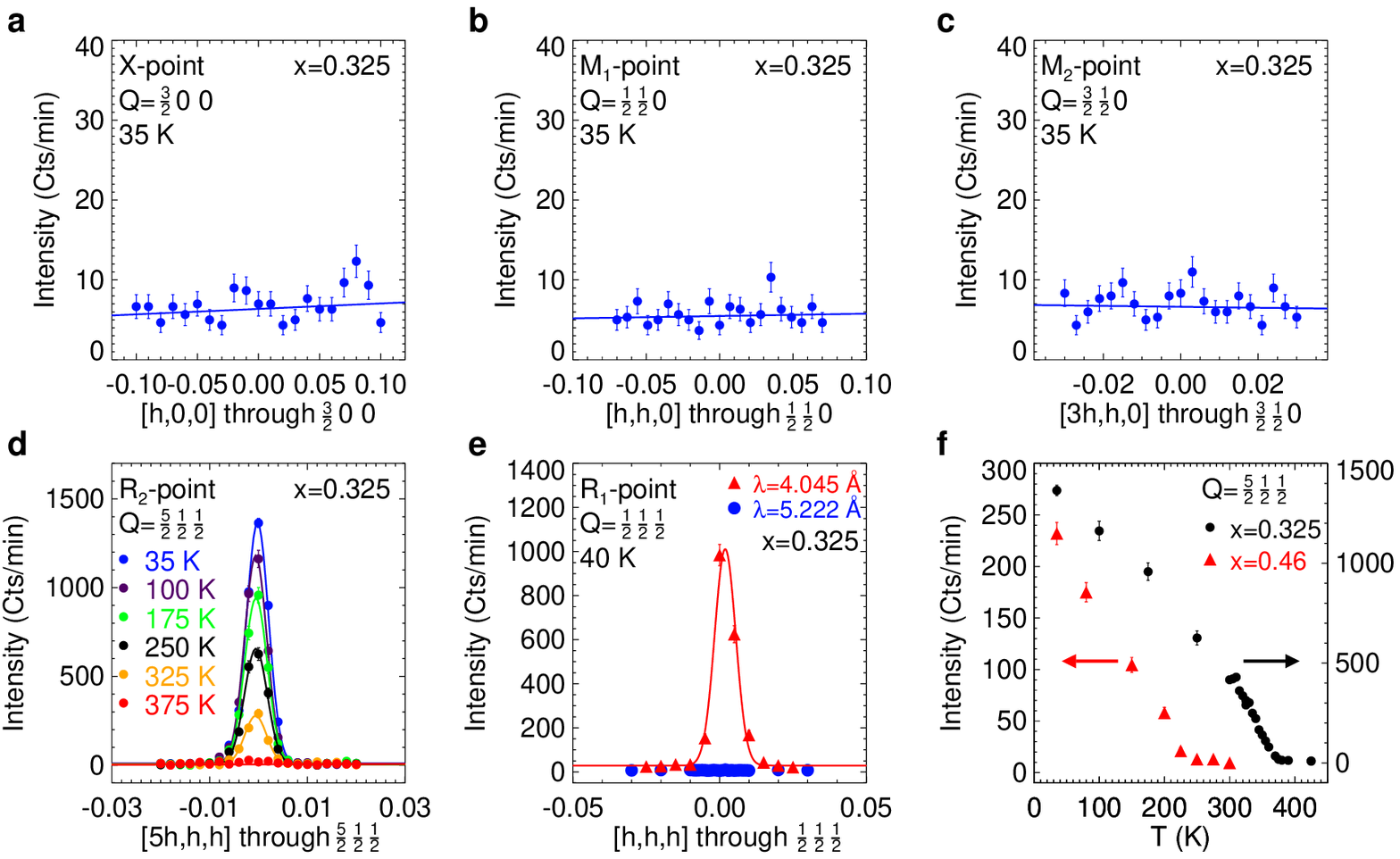}
\clearpage
\includegraphics[scale=0.9]{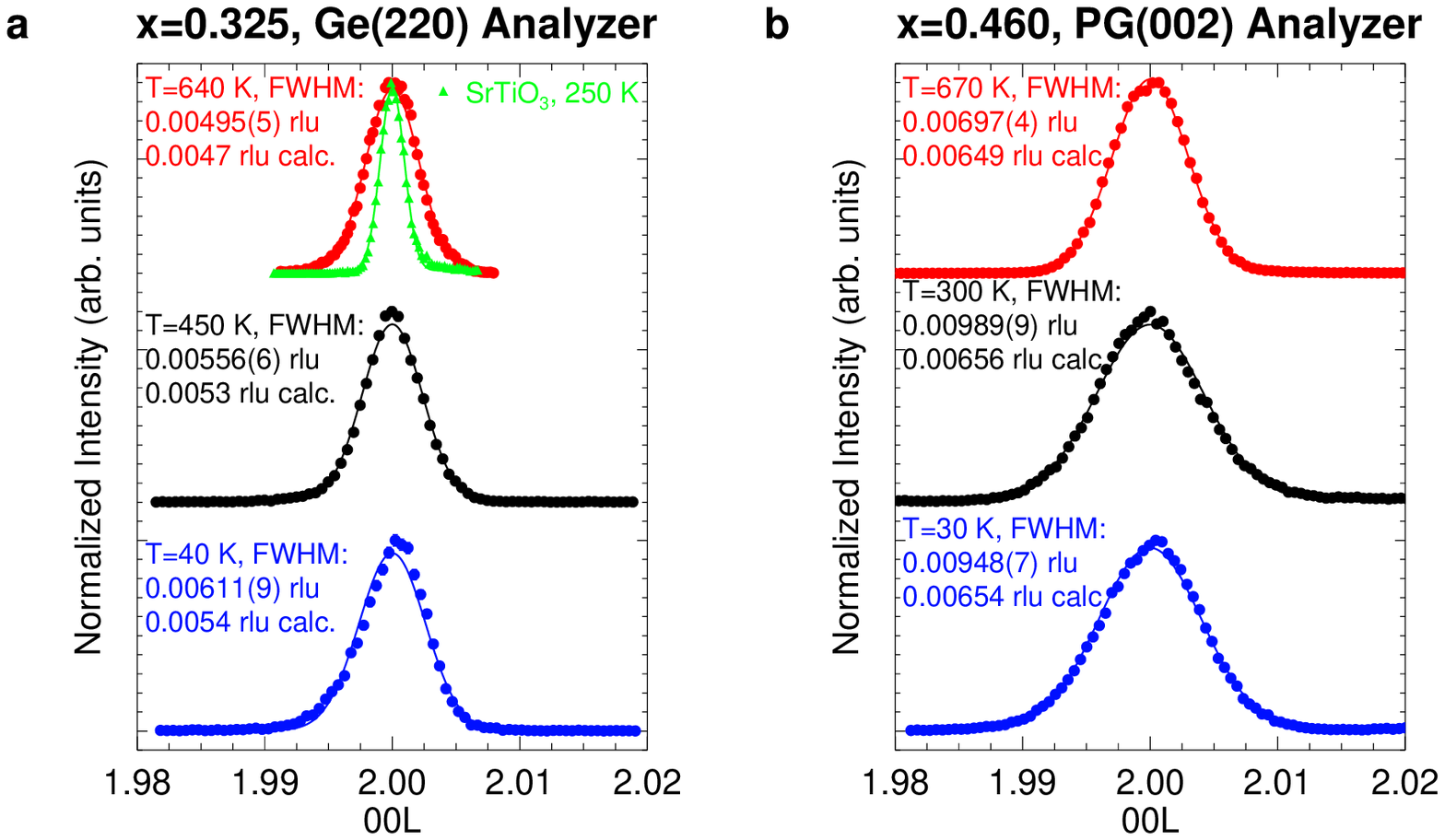}
\clearpage
\includegraphics[scale=1.0]{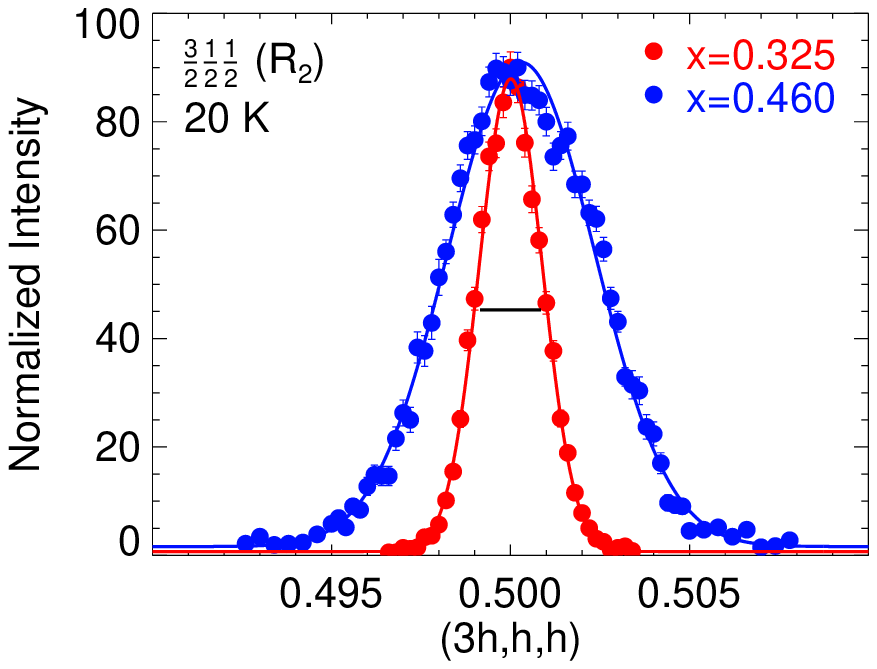}

\end{document}